\documentstyle[12pt]{article}

\begin{document}
\def\la{\mathrel{\mathpalette\fun <}}
\def\ga{\mathrel{\mathpalette\fun >}}
\def\fun#1#2{\lower3.6pt\vbox{\baselineskip0pt\lineskip.9pt
        \ialign{$\mathsurround=0pt#1\hfill##\hfil$\crcr#2\crcr\sim\crcr}}}
\newcommand {\eegg}{e^+e^-\gamma\gamma~+\not \! \!{E}_T}
\newcommand {\mumugg}{\mu^+\mu^-\gamma\gamma~+\not \! \!{E}_T}
\renewcommand{\thefootnote}{\fnsymbol{footnote}}
\bibliographystyle{unsrt}

\begin{flushright}
UMD-PP-99-20\\
\end{flushright} 
\begin{center}
{\Large \bf Bimaximal Neutrino Mixing and Neutrino Mass Matrix}

\vskip0.8cm

{\bf  Rabindra N. Mohapatra$^1$, and Shmuel Nussinov$^2$}\\

\vskip1.0cm

{\it$^{(1)}${ Department of Physics, University of Maryland,
College Park, Md-20742, USA.}}

{\it$^{(2)}${ Department of Physics, Tel Aviv University, Tel Aviv,
Israel\\
and Department of Physics and Astronomy, Univ. of South Carolina,
Columbia, SC 29208, USA
}}

\end{center}

\begin{abstract}
 We show that the bimaximal neutrino mixing pattern suggested by the 
solar and atmospheric neutrino data 
can be derived from the maximal,  symmetric, four neutrino
 mixing in the limit that
one of the neutrinos is made heavy. Imposing the constraints of no 
neutrinoless double beta decay and a 20\% hot dark matter component of 
the universe leads to a three neutrino mass matrix recently suggested 
by Georgi and Glashow. Our result can be useful in constructing
theoretical models for the bimaximal pattern. We illustrate this by a simple 
example.
 \end{abstract}

\newpage
\renewcommand{\thefootnote}{\arabic{footnote})}
\setcounter{footnote}{0}
\addtocounter{page}{-1}
\baselineskip=24pt

 \vskip0.5cm
%\section{Introduction}

Recent observation of neutrino oscillations at Super-Kamiokande $^{[1]}$ has
been a great source of excitement in particle physics for two reasons: first
it provides the best evidence to date for  
nonzero neutrino mass which, in turn, is the first
sign for new physics 
beyond the standard model; the second reason is that the 
Super-K analysis seems 
to require maximal mixing between  $\nu_{\mu}$ and $\nu_{\tau}$.
The latter mixing pattern is very different from that observed in the quark 
sector suggesting  a more fundamental distinction between the
physics of neutrino and the charged fermion masses. Note that ultralightness 
of the neutrinos is another such distinction and seems to require the 
novel phenomena of seesaw mechanism to understand it. 

The apparent maximal mixing between the $\nu_{\mu}$ and $\nu_{\tau}$ has
led to speculations that the entire mixing pattern in the three
neutrino sector may be essentially maximal.
In this case, the solar neutrino puzzle will be solved 
either by the vacuum or large angle MSW mechanism\cite{bahcall}. The hot 
dark matter of the universe which may require a total neutrino mass in the 
range of 4 to 5 eV will then imply that all three neutrino species are 
degenerate in mass\cite{caldwell}. A crucial test of this model will
also be provided by further improvement of the solar neutrino data e.g.
day-night variation will test the large angle MSW mechanism which is one 
possibility in this case. The incorporation of the LSND data\cite{lsnd}
seems to require a fourth (sterile) neutrino, though there have been 
attempts to fit all data within a three neutrino, roughly maximal, mixing 
framework, using ``indirect neutrino oscillations''. For a recent version,
 see e.g. \cite{thun}.

Three  
possible neutrino mass matrices $U_{\nu}$ defined via
\begin{eqnarray}
\left(\begin{array}{c} \nu_e\\ \nu_{\mu} \\ \nu_{\tau}\end{array}
\right)=~ U_{\nu}\left(\begin{array}{c} \nu_1 \\ \nu_2 \\ \nu_3
\end{array} \right)\end{eqnarray}
have been widely discussed in the literature:
\noindent{\it Case (A)\cite{fritzsch}:}
\begin{eqnarray}
U_{\nu}=\left(\begin{array}{ccc}
\frac{1}{\sqrt{2}} & -\frac{1}{\sqrt{2}} & 0\\
\frac{1}{\sqrt{6}} &\frac{1}{\sqrt{6}} &-\frac{2}{\sqrt{6}} \\
\frac{1}{\sqrt{3}} &\frac{1}{\sqrt{3}} &\frac{1}{\sqrt{3}}
\end{array}\right)
\end{eqnarray}
We will call this the democratic mixing.

\noindent{\it Case (B)\cite{gold}:}
\begin{eqnarray}
U_{\nu}=\left(\begin{array}{ccc}
\frac{1}{\sqrt{2}} & -\frac{1}{\sqrt{2}} & 0\\
\frac{1}{2} &\frac{1}{2} &\frac{1}{\sqrt{2}} \\
\frac{1}{2} &\frac{1}{2} &-\frac{1}{\sqrt{2}}
\end{array}\right)
\end{eqnarray}
This has been called in the literature bimaximal mixing\cite{gold}.

\noindent{\it Case (C)\cite{wolfenstein}:}
\begin{eqnarray}
U_{\nu}=\frac{1}{\sqrt{3}} \left(\begin{array}{ccc}
1 & \omega & \omega^2\\
1 & \omega^2 &\omega\\
1 & 1 & 1 \end{array} \right)
\end{eqnarray}
where $\omega =e^{\frac{2\pi i}{3}}$; we will call this the maximal 
symmetric mixing. Such a mixing matrix $U_{(n)}$ emerges when we impose 
maximal $(\frac{1}{n})$ suppresion of each neutrino
flavor\cite{nussinov} for any prime $n$. If $n = p_{1}^{m_{1}} \ldots p_{k}^{m_{k}}$,
where $p_{i}$ are prime integers, then the Maximal Mixing Symmetric (MMS)
matrix will consist of a direct product of $m_{1}$ MMS matrices corresponding
to $p_{1}$ with $m_{2}$ MMS matrices corresponding to $p_{2}$ etc.

$U_{(3)}$ (case (C) may be marginal if we take the CHOOZ\cite{chooz}
and the atmospheric neutrino data into account. The cases (A) and (B)
are however fully consistent with the CHOOZ and all other data as long as
LSND data is not included.

Should any of these mixing patterns be confirmed by further data, a key 
theoretical challenge would be to understand them from a fundamental gauge
theory framework. Clearly, one must look for some underlying symmetry
that would lead to the entries in the above mixing matrices. 
In a previous note we showed that the democratic mixing matrix\cite{moh} 
can be understood if the lepton sector of the gauge model has a dicrete
interchange symmetry among three generations. No such simple symmetry among 
three generations of leptons is apparent for the bimaximal case (A). 
It is our goal in this paper to show that the bimaximal pattern emerges
from a symmetric $4\times 4$ maximal mixing matrix if a particular linear 
combination of the neutrinos is made heavy and decouples.
Needless to say that LEP 
and SLC data would imply that if there is a fourth generation, the 
neutrinos of the fourth generation must be heavy enough not to 
contribute to the Z-width measured at LEP and SLC. Thus its decoupling 
from light neutrino mass matrix is to be expected on phenomenological 
grounds. This result may provide us a clue regarding how one can proceed in 
constructing a gauge model for the bimaximal pattern. In particular, it
highlights the need for the existence of a fourth generation for this 
purpose. We give an example of such a construction. 

To introduce our line of reasoning, we note that
for two neutrino flavors any mass matrix of 
the form:
\begin{eqnarray}
M_2=\left(\begin{array}{cc}
a & b\\
b & a\end{array} \right)
\end{eqnarray}
leads to the maximal mixing matrix of the form
$\frac{1}{\sqrt{2}}\left(\begin{array}{cc}1 & 1 \\
1 & -1 \end{array}\right) \equiv U_{2}$.

The unique maximal mixing $4 \times 4$ mixing matrix is then 
given by the direct product 

\begin{eqnarray}
U_{4} = U_{2} \otimes U_{2} = \frac{1}{2} \left(\begin{array}{cccc} 1 & 1 & 1 & 1\\
1 & -1 & 1 & -1 \\ 1 & 1 & -1 & -1 \\ 1 & -1 & -1 & 1 \end{array}\right).
\end{eqnarray}
$U_{4}$ arises naturally in a diagonalization of a $4 \times 4$ mass 
matrix of the form: $M_{4} = M_{2} \otimes M_{2}^{\prime}$;
$M_{2}^{\prime}= \left(\begin{array}{cc} c & d \\ d & c\end{array} \right).$
In the $(\nu^0_e, \nu^0_\mu, \nu^0_\tau, \nu^0_E)$ basis:
\begin{eqnarray}
M_4 =\left(\begin{array}{cccc} A & B & C & D \\ B & A & D & C\\ C & D & A & 
B \\ D & C & B & A \end{array}\right),
\end{eqnarray}
with $A \equiv ac$, etc. 
Let us now add to the four neutrino mass Lagrangian a heavy mass for the 
combination $\nu^0_e -\nu^0_E$
\begin{eqnarray}
{\cal L} = M(\nu^0_e-\nu^0_E)(\nu^0_e-\nu^0_E)
\end{eqnarray}
Calling $\nu_\pm\equiv (\nu^0_e \pm \nu^0_E)/\sqrt{2}$, we can rewrite the 
above 
mass matrix in terms of $\nu_+, \nu_\mu, \nu_\tau, \nu_-$ and decouple the
field $\nu_-$. Identifying $\nu_+=\nu_e$, we get for the $\nu_e,\nu_\mu, 
\nu_\tau$ mass matrix:
\begin{eqnarray}
M_3 = \left(\begin{array}{ccc} A+D & F & F\\ F & A & D \\ F & D & A
\end{array}\right)
\end{eqnarray}
where $F\equiv (C+B)/\sqrt{2}$. Diagonalizing this 
$3\times 3$ mass matrix $M_3$, we find that the 
neutrino mixing matrix corresponds to the bimaximal case.
This proves the main assertion of our model.

The eigenvalues are
$m_1 = A+D+\sqrt{2} F; m_2 = A+D-\sqrt{2} F; m_3 = A-D$. We can accomodate 
a hot dark matter in this model if all three masses are almost degenerate. 
Combining this with the requirement that the model satisfy the 
neutrinoless double beta decay constraint\cite{klap} leads to the following
ordering of the parameters:
\begin{eqnarray}
A+D \simeq \delta_2: F\simeq \sqrt{2} A +\delta_1
\end{eqnarray}
where the~ S.K.~atmospheric neutrino fit requires that 
$\delta_1\simeq 10^{-3}$ eV (assuming that overall common mass for the 
neutrinos is $\simeq 1-2$ eV). Similarly, the solution to the solar 
neutrino puzzle via the large angle MSW solution requires that 
$\delta\simeq 10^{-4}-10^{-5}$ eV.

It is interesting to note that after enforcing the neutrinoless double        
beta decay and the hot dark matter constraints, $\delta_1\to 0$
$\delta_2\to 0$, we obtain the 
following neutrino mass matrix:                                                     
\begin{eqnarray}                                                          
M_{\nu}= m_0\left(\begin{array}{ccc} 0 & \frac{1}{\sqrt{2}} & 
\frac{1}{\sqrt{2}} \\ \frac{1}{\sqrt{2}} & \frac{1}{2} & -\frac{1}{2}\\
\frac{1}{\sqrt{2}} & -\frac{1}{2}  & \frac{1}{2} \end{array}\right)
\end{eqnarray}
This is precisely the mass matrix suggested by Georgi and 
Glashow\cite{georgi} as a possible way to accomodate the neutrino data.
 We therefore have an alternative derivation of 
the same mass matrix.

Let us briefly mention possible implications of our result. In addition to
hinting a deeper structure behind the bimaximal pattern,
our result may 
have practical applications for building gauge models for this pattern. One
avenue to explore would be to consider a four generation version of the 
standard model\cite{hung}
and decouple the heaviest Majorana combination as discussed above to 
obtain the three generation model with bimaximal mixing. 
Below, we give an example of such a construction.

\noindent\underline{{\it A four generation model for bimaximal mixing:}}

Consider a four generation extension of the standard model. Let us denote the
lepton doublets by $L_i$ with $i=1,2,3,4$ and the four right handed 
charged leptons by $e^c_i$. Let us add two right handed 
($SU(2)_L$ singlet) neutral fermions denoted by $(\nu^c_1,\nu^c_2)$. 
We omit the quark sector for simplicity. We also 
augment the Higgs sector of the model by adding an iso-triplet $Y=2$ field
$(\Delta)$, whose neutral component acquires a vev which is seesaw 
suppressed to be of order of eV's. We then add four new downlike Higgs 
doublets $H_a$ ($a=1,2,3,4$) and two new uplike Higgs pair in addition to 
the two standard model like doublets $\phi_a$ (a=1,2). The up and down like 
Higgs doublets 
can distinguished from each other by a Peccei-Quinn like symmetry (or 
supersymmetry) which 
we do not display since it can be softly broken. To extract the key 
features of our idea, we impose a discrete symmetry
$D_1\times D_2$ on our model. Under this symmetry, we assume the fields 
to transform as in Table I: 

\begin{center}
Table I
\end{center}
\begin{center}
\begin{tabular}{|c||c|}\hline
$D_1 $& $D_2$ \\ \hline
$L_1\leftrightarrow L_2$ & $L_1\leftrightarrow L_3$\\
$ L_3\leftrightarrow L_4$ & $L_2\leftrightarrow L_4$ \\
$\nu^c_1\leftrightarrow \nu^c_2$ & $\nu^c_1\leftrightarrow -\nu^c_2$\\
$ \phi_1\leftrightarrow \phi_2$ &$\phi_1\leftrightarrow \phi_2$ \\
$H_1\leftrightarrow H_2$ & $ H_1\leftrightarrow -H_2$\\
$H_3\leftrightarrow H_4$ & $H_3 \leftrightarrow H_4$\\
$e^c_{3,4}\leftrightarrow e^c_{3,4}$ & $ e^c_{3,4}\leftrightarrow 
-e^c_{3,4}$\\ \hline \end{tabular}
\end{center}
\noindent {Table caption: Transformation properties of the fields under 
the interchange symmetries $D_1$ and $D_2$.}

 The rest of the fields are all assumed to be singlets under these 
symmetries. Let us now write down the invariant Yukawa couplings under 
these symmetries, ${\cal L} ={\cal L}_1 +{\cal L}_2 +{\cal L}_3 $
\begin{eqnarray}
 {\cal L}_1= f_1 (L_1 L_1 +L_2 L_2 + L_3 L_3 +L_4 L_4 )\Delta \nonumber\\
+f_2 (L_1 L_4+L_2 L_3)\Delta \nonumber\\
+f_3(L_1 L_3+L_2 L_4)\Delta\nonumber\\
+f_4(L_1 L_2 +L_3 L_4)\Delta + h.c.
\end{eqnarray}

Note that after the $\Delta^0$ field acquires a seesaw suppressed vev, it 
gives rise to the mass matrix $M_4$ in Eq. 7. Now we have to show that 
consistent with the discrete symmetry, we can decouple the combination 
of neutrinos in Eq. 8 and  
that the charged lepton mass matrices are such that they do not 
induce any contributions to mixings. To see the second point,
 let us write down $\cal 
L_2$.
\begin{eqnarray} 
{\cal L}_2=h_1 [ (L_1-L_4)H_3 + (L_2-L_3)H_4] e^c_3\nonumber\\
+ h_2 [(L_1+L_4)H_1 +(L_2 +L_3)H_2]e^c_4]\nonumber\\
+h_3 [ (L_2 +L_3 )H_3 + (L_1 +L_4)H_4 ](e^c_2+\alpha e^c_1)\nonumber\\
+h_4 [ (L_2 -L_3)H_1 +(L_1-L_4)H_2 ](e^c_1 +\beta e^c_2) + h.c.
\end{eqnarray}
We assume that $h_1\gg h_3 \gg h_4 \gg h_2 $.
If we assume that only $H_{1,3}$ have vev then it keeps the $e_1-e_4$
as a separate eigen state, which we will assume to be heavy and make it 
decouple. The $e_1+e_4$ will be the lightest eigenstate to be identified 
with the electron state.

Next let us show how one can decouple the correct heavy neutrino 
combination so as to implement 
our suggestion. For this purpose we note that there is the followinf Yukawa
coupling allowed by the symmetries of our model:
\begin{eqnarray}
{\cal L}_3= f_5 [(L_1-L_4)\phi_1\nu^c_1+(L_2-L_3)\phi_2\nu^c_2]\nonumber\\
+f_6 [(L_1-L_4)\phi_2\nu^c_1+(L_2-L_3)\phi_1\nu^c_2]
\end{eqnarray}
We add a soft breaking term to the theory of the form $M\nu^c_2\nu^c_2$ with
$M\simeq 10^{11}$ GeV or so and assume that $f_6\ll f_5$ and 
$<\phi_1>\neq 0$. This then leads to 
a partial seesaw that makes one of the light neutrinos i.e. $\nu_2-\nu_3$
very light whereas it leaves the othet combination  i.e. $\nu_1-\nu_4$ with
a mass in the 100 GeV range (not seesaw suppressed). As a result this 
combination decouples from our low energy neutrino spectrum as discussed.
The three low energy neutrinos in our model have bimaximal mixing pattern.
We thus see that our decoupling scheme can be realized in explicit gauge 
models. We realize that the model just presented, though realistic
is given in the spirit of providing an existence proof of
the kind of schemes we are advocating. However, they can perhaps be 
simplified and embedded into more clever schemes that will then provide
a deeper origin for bimaximal models. 

In conclusion, we have suggested a way to construct gauge models for 
bimaximal neutrino mixing starting with a four generation model and
illustrate this with the help of an extension of the standard model that 
incorporates four neutrino generations.

 The work of R. N. M. is supported by the
National Science foundation under grant PHY-9802551.
Both of us would like to acknowledge the US-Israeli Binational 
Science Foundation grant.

\end{document}